\documentstyle[11pt,fleqn,a4]{article}
\def\co{\Delta}
\def\coe{\varepsilon}
\def\ps{{\cal P}_S}
\def\psind#1#2#3#4{{{\ps}^{#1#2}}_{#3#4}}
\def\paind#1#2#3#4{{{\pa}^{#1#2}}_{#3#4}}
\def\ptind#1#2#3#4{{{\pt}^{#1#2}}_{#3#4}}
\def\bfpaind#1#2#3#4{{{\mbox{\boldmath $ \pa $}}^{#1#2}}_{#3#4}}

\def\bfptind#1#2#3#4{{{\mbox{\boldmath $ \pt $}}^{#1#2}}_{#3#4}}

\def\pa{{\cal P}_A}
\def\pt{{\cal P}_1}
\def\RM{\hat R}

\def\bfRM{\hat {\bf R}}

\def\bfRmvier#1#2#3#4{{\bf\RM}^{#1#2}{}_{#3#4}}
\def\bfRmivier#1#2#3#4{{\bf\RM}^{-1{}#1#2}{}_{#3#4}}

\def\Rm#1#2#3#4{{\RM}^{#1#2}{}_{#3#4}}
\def\Rmi#1#2#3#4{{\RM}^{-1{}#1#2}{}_{#3#4}}

\def\t#1#2{{t^{#1}}_{#2}}
\def\th#1#2{{{\hat{t}}^{#1}}{}_{#2}}
\def\C#1#2{C_{#1#2}}
\def\Ci#1#2{C^{#1#2}}
\def\Yps#1#2{{Y^{#1}}_{#2}}
\def\Ypsh#1#2{{{\hat Y}^{#1}}_{#2}}

\def\Lplus#1#2{{{L^+{}}^{#1}}_{#2}}
\def\Lminus#1#2{{{L^-{}}^{#1}}_{#2}}
\def\Lplusmin#1#2{{{L^{\pm{}}}^{#1}}_{#2}}
\def\Lh{\hat L}
\def\Lhplus#1#2{{\Lh}^{+{}#1}{}_{#2}}
\def\Lhminus#1#2{{\Lh}^{-{}#1}{}_{#2}}
\def\Lhplusmin#1#2{{\Lh}^{\pm{}#1}{}_{#2}}

\def\qzahl#1#2{{[#1]}_{#2}}
\def\aqzahl#1#2{{\{#1\}}_{#2}}
\def\q2{{[2]}_q}

\def\l{\lambda}
\def\csind#1#2{c^{\prime}_{\pm,#1,#2}}
\def\cind#1#2{c_{\pm,#1,#2}}
\def\dob#1{{\partial}^{#1}}
\def\dobb#1{{\bar{{\partial}}}^{#1}}

\def\ot{\otimes}

\def\ket#1#2#3#4{|#1,#2,#3,#4\rangle}
\def\ketpm#1#2#3#4{|\pm,#1,#2,#3,#4\rangle}

\def\cket#1#2#3#4{#1,#2,#3,#4\rangle}
\def\cketpm#1#2#3#4{\pm,#1,#2,#3,#4\rangle}
\def\bra#1#2#3#4{\langle #1,#2,#3,#4|}
\def\brapm#1#2#3#4{\langle \pm,#1,#2,#3,#4|}
\def\regfunlor{U_{\cal R}}

\def\mq{A_x({\cal M}_{\lowercase{q}})}
\def\mqp{A_p({\cal M}_{\lowercase{q}})}
\def\Adreh{A_x ({\cal U}_q^3,{\cal M}_q)}
\def\Apdreh{A_p ({\cal U}_q^3,{\cal M}_q)}
\def\Alor{A_x({\cal U}_q,{\cal M}_q)}
\def\Aplor{A_p({\cal U}_q,{\cal M}_q)}
\def\Ahutlor{A_x({\hat{{\cal U}}}_q,{\cal M}_q)}
\def\Alormu{A_{x,\mu}({\cal U}_q,{\cal M}_q)}
\def\Aplormu{A_{p,\mu}({\cal U}_q,{\cal M}_q)}
\def\Ahutlormu{A_{x,\mu}({\hat{{\cal U}}}_q,{\cal M}_q)}
\def\Udreh{{\cal U}_{\lowercase{q}}^3}
\def\Ulor{{\cal U}_{\lowercase{q}}}
\def\Uhutlor{{\hat{{\cal U}}}_q}

\def\difflor{D({\cal M}_q)}

\def\h{{\cal H}}
\def\qi{q^{-1}}
\def\vepl{V_1^+}
\def\vemi{V_1^-}
\def\vzpl{V_2^+}
\def\vzmi{V_2^-}
\def\Menge#1{\ifmmode
           {\rm I\mkern -3mu #1}%  18mu = 1em
         \else
           {\rm I\kern -0.15em #1}%
         \fi}

\def\Mengescript#1#2#3{
         {\rm\setbox1=\vbox{#1}\vrule height 0.66\ht1 width #2 depth 0pt%
         \kern #3 #1 }%
         }

\def\N{\Menge{N}}
\def\R{\Menge{R}}
\def\Z{Z\hspace*{-1.2em}Z}

\def\ce{{\rm\setbox1=\vbox{C}\hskip 2.5pt
         \vrule height 0.95\ht1 width 0.7pt depth -0.5pt%
         \kern -3.7pt C}%
	 }
\def\Ce{{\rm\setbox1=\vbox{C}\hskip 2.5pt
         \vrule height 0.95\ht1 width 0.7pt depth -0.5pt%
	 \kern -5.2pt C}%
	 }
\def\CE{{\rm\setbox1=\vbox{C}\hskip 2.5pt
         \vrule height 0.95\ht1 width 0.7pt depth -0.5pt%
	 \kern -7.5pt C}%
	 }
\def\scriptC{{\rm\setbox1=\vbox{C}\hskip 2.5pt
         \vrule height 0.66\ht1 width 0.5pt depth -0.45pt%
         \kern -2.75pt C}%
         }
\newtheorem{defin}{Definition}[section]
\newtheorem{lem}[defin]{Lemma}
\newtheorem{prop}[defin]{Proposition}

\newcommand{\bea}{\begin{eqnarray}}
\newcommand{\eea}{\end{eqnarray}}
\newcommand{\be}{\begin{equation}}
\newcommand{\ee}{\end{equation}}
\newcommand{\ra}{\rightarrow}
\newcommand{\as}[1]{\mbox{${#1}^{\ast}$}}
\def\thebibliography#1{\section*{References}
  \list
  {[\arabic{enumi}]}{\settowidth\labelwidth{[#1]}
  \leftmargin\labelwidth
  \advance\labelsep8pt
  \advance\leftmargin\labelsep\parsep0.0em\itemsep0.0em
  \usecounter{enumi}} \relax}

\begin{document}

% ------------------------------------------------------------
% Falls alle Eintraege ueber BIBTEX in das Literaturverzeichnis solen
% \nocite{*}

% ------------------------------------------------------------
% Tabelle/File mit Trennhilfen

% ------------------------------------------------------------
% Titelblatt
\thispagestyle{empty}
\title{\begin{flushright} {\normalsize LMU-TPW 94-4\\ August, 1994}
\end{flushright}
\vspace{3cm} Hilbert Space Representation of an Algebra of Observables for
q-Deformed Relativistic Quantum Mechanics}
\author{W. Zippold}
\date{Sektion Physik, Universit\"at M\"unchen\\
Theresienstr. 37, D-80333 Munich, Germany}
\maketitle

\begin{abstract}
Using a representation of the q-deformed Lorentz algebra as differential
operators on quantum Minkowski space, we define an algebra of
observables for a q-deformed relativistic quantum mechanics with spin zero.
We construct a Hilbert space representation of this algebra in which the
square of the mass $ p^2 $ is diagonal.
\end{abstract}

% Zusammenfassung

% ------------------------------------------------------------
% Inhaltsverzeichnis

%\tableofcontents
%\addcontentsline{toc}{chapter}{Einf"uhrung}
% ------------------------------------------------------------
% Normale Texte

% Einf"uhrung -----------------------------
\section{Introduction}
The concept of Lie groups and Lie algebras has found a natural generalization
in the framework of non-commutative Hopf algebras and quantum groups. This has
posed the question,
whether these can appear as symmetries of physical theories. In this paper we
want to address the problem of quantum symmetric relativistic quantum mechanics
with spin zero.

In ordinary relativistic quantum mechanics, we have a Hilbert space of
states which is a representation of the Poincar\'{e} algebra. Acting on this
space, we have the algebra of observables generated by position and momentum
coordinates, which are essentially self-adjoint operators defined on a common
domain, which is dense in the Hilbert space. They satisfy the Heisenberg
algebra
$ [x^i,p^j] = ig^{ij} $. Position and momentum space are both isomorphic
to Minkowski space and carry a representation of the Lorentz group.

In this paper, we define a deformation of the algebra of observables. The
position and momentum coordinates generate algebras, which are isomorphic
as $ \ast $-comodule algebras of the q-deformed Lorentz group. Introducing an
algebra of angular momentum operators, which is a representation $
U_q(SL(2,\ce) $,
we construct a Hilbert space representation of this algebra of observables.
Coordinates and momenta are represented on this Hilbert space by unbounded
operators.

% kapitel 1 -------------------------------
\section{Preliminaries}
\subsection{$ SL_q(2,\Ce) $ and Quantum Lorentz Group}
The complex quantum group $ SL_q(2,\ce) $ (\cite{dswz}) is a $ \ast $-Hopf
algebra constructed by taking two copies
of the quantum group $ SL_q(2) $ (cf. \cite{FRT}) with
generators $ (\t {\alpha}{\beta})_{\alpha,\beta = 1,2} $ and
$ (\th {\alpha}{\beta})_{\alpha,\beta = 1,2} $ which are connected by the mixed
relations $ \Rm {\alpha}{\beta}{\gamma}{\delta}\th{\gamma}{\rho}\t{\delta}
{\sigma} = \t{\alpha}{\mu}\th{\beta}{\nu}\Rm{\mu}{\nu}{\rho}{\sigma} $, where
$ \RM $ denotes the $ \RM $-matrix of $ SL_q(2) $. The involution is
$ (\t{\alpha}{\beta})^{\ast} := S(\th{\alpha}{\beta}) $,
$ S $ being the antipode of $ SL_q(2) $.

The quantum Lorentz group $ SO_q(3,1) $ (cf. \cite{cssw}) is the
$ \ast $-sub-Hopf algebra
$ \subset SL_q(2,\ce) $ generated by the elements $ M^{(\alpha\beta)}{}_
{(\gamma\delta)} :=
\th{\alpha}{\gamma}\t{\beta}{\delta} $ with relations and Hopf structure
induced by $ SL_q(2,\ce) $. Its $ \bfRM $-matrix is a product of
$ SL_q(2) $-$ \RM $-matrices.
Using a single index $ i= 1,2,3,4 $ instead of the double
index $ (\alpha,\beta) = (1,1),(1,2),(2,1),(2,2) $, the reality condition for
the generators is given by $ {(M^i{}_j)}^{\ast} = \eta^j{}_b S(M^b{}_a)
\eta^a{}_i $, where the matrix $ \eta^i{}_j $ is essentially given by the
$ \RM $-matrix of $ SL_q(2) $.The $ \bfRM $-matrix has the projector
decomposition characteristic of orthogonal quantum groups:
\be
\bfRmvier ijkl = q\psind ijkl -\qi\paind ijkl + q^{-3}\ptind ijkl,
\ee
$ \ps $, $ \pa $ and $ \pt $ being the symmetric, antisymmetric and trace
projector, respectively. $ \bfptind ijkl $ can be expressed in terms of the
q-Minkowski metric $ \C ij $ as
$ \bfptind ijkl = Q^{-1}\Ci ij\C kl $ with $ Q := \Ci ij\C ij = \q2^2 $. Here
we have introduced the q-numbers $ \qzahl xq := (q^x-q^{-x})/(q-\qi) $.

\subsection{Quantum Minkowski Space and Differential Calculus}
A comodule algebra for $ SO_q(3,1) $ can be defined as the unital \ce-algebra
generated by the coordinates $ x^i $ with relations $ \bfpaind ijkl x^k x^l = 0
$.
This is the algebra of functions on quantum Minkowski space and is denoted
$ \mq $. The coaction
$ \delta $ of $ SO_q(3,1) $ is given by $ \delta(x^i) := M^i{}_j \ot x^j $. For
$ \delta $ to become a homomorphism of $ \ast $-algebras, we have to define
the involution on $ \mq $
by $ (x^i)^{\ast} := x^k\C kl\eta^l{}_i $. The element $ r^2 := \C ij x^ix^j $
is central in $ \mq $. Another convenient set of generators
$ t,x,y,z $ for $ \mq $ is given by \\
\parbox{6.0cm}{\bea
t & := & \qi/\q2(\qi x^2 - x^3),\nonumber\\
y & := & x^1,\nonumber
\eea}\hfill
\parbox{6.0cm}{\bea
z & := & \qi/\q2(-q x^2 - x^3),\\
x & := & -x^4.\nonumber
\eea}\\
The generator $ t $ becomes central in $ \mq $ and factorization with respect
to the relation $ t=0 $ yields an $ SO_q(3) $-comodule algebra (cf.
\cite{cssw}).

We can now introduce partial derivatives acting on $ \mq $
as linear operators. We define the partial derivative $ \dob i\in
{\mbox{End}}_{\scriptC}(\mq),i=1,\ldots,4, $ by its action
$ \dob i ({\bf1}) := 0 $ on the
unit element $ {\bf 1}\in \mq $ and by the Leibniz rule
\be
\dob i(x^jf) := \Ci ij f + q\Rmi ijkl x^k\dob l(f)\qquad\forall f\in \mq,
\ee
\begin{sloppypar}which determines $ \dob i $ on the whole algebra $ \mq $. The
partial derivatives satisfy $ \bfpaind ijkl\dob k\dob l=0 $ and
$ \delta (\dob i(f)) = (M^i{}_j\ot\dob j)\circ\delta(f),\;\;
\forall f\in \mq $.\end{sloppypar} Therefore the algebra $ A_{\partial} $
generated by
$ (\dob i) $ is isomorphic to $ \mq $ as an $ SO_q(3,1) $-comodule algebra.
The coordinates $ x^i $ act on $ \mq $ as linear
operators by left multiplication. So we can consider the
algebra \\
$ A_{x,\partial} := <(x^i)_{i=1,\ldots,4},(\dob i)_{i=1,\ldots,4}>
\subset {\mbox{End}}_{\scriptC}(\mq) $ with relations
\bea
\bfpaind ijkl x^k x^l & = & \bfpaind ijkl \dob k\dob l =0,\nonumber\\
\dob i x^j & = & \Ci ij + q\Rmi ijkl x^k\dob l.
\eea
Given this
algebra, we can recover the Leibniz rule and therefore the action of $ \dob i $
as a differential operator on $ \mq $. Next, we want to define a $ \ast $-
structure on $ A_{x,\partial} $. To that end, we first note, that there exists
an operator $ \Lambda\in A_{x,\partial} $ (cf. \cite{ogi}) satisfying
$ \Lambda({\bf 1}) = {\bf 1} $, $ \Lambda x^i = q^2 x^i\Lambda\; $ and
$ \Lambda\dob i = q^{-2}\dob i\Lambda $. Consequently $ \Lambda\in{\mbox{End}}_
{\scriptC}(\mq) $ is a positive and invertible operator, so we can define
an invertible operator $ \mu := q^2\Lambda^{1/2} $. We can now introduce the
algebra $ \difflor $ generated by the coordinates $ (x^i) $,
the derivatives $ (\dob i) $ and the operators $ \mu $ and $ \mu^{-1} $, and we
call it {\em algebra
of differential operators} on quantum Minkowski space. In this extended algebra
we introduce the elements $ \dobb i $ by
\be
\dobb i := \Lambda^{-1}(\dob i + q^3\l x^i\Delta),
\ee
and it was shown in \cite{ogi} that these differential operators satisfy the
relations of the second possible covariant differential calculus on $ \mq $,
i.e. we have $ \bfpaind ijkl\dobb k\dobb l = 0 $ and $ \dobb i x^j = \Ci ij +
\qi\Rm ijkl x^k\dobb l $. The involution on the partial derivatives can now
be defined by $ {(\dob i)}^{\ast} := - q^{-4}\dobb k\C kl\eta^l{}_i $, such
that $ \difflor $ finally becomes a $ \ast $-algebra in which $ \mu^{\ast} =
\mu^{-1} $.

\subsection{Regular Functionals and Vector Fields}

In \cite{dswz} it was shown that the dual algebra $ {SL_q(2,\ce)}^{\ast} $ of
$ SL_q(2,\ce) $ contains a $ \ast $-sub-Hopf algebra $ \regfunlor $, called
{\em algebra of regular functionals}. $ \regfunlor $ is generated by
functionals $ {(\Lplusmin {\alpha}{\beta})}_{\alpha,\beta =1,2} $ and
$ {(\Lhplusmin {\alpha}{\beta})}_{\alpha,\beta =1,2} $. The action of these
functionals is defined using the $ \RM $-matrix of $ SL_q(2) $:
\be
\Lplusmin {\alpha}{\beta}({\bf 1}) := \delta^{\alpha}{}_{\beta}\qquad
\Lplusmin {\alpha}{\beta} (\t {\gamma}{\delta}) := {\RM}^{\pm 1}{}^{\alpha
\beta}{}_{\delta\gamma}\qquad \Lplusmin {\alpha}{\beta} (\th {\gamma}{\delta})
:= {\RM}^{\mp 1}{}^{\alpha\beta}{}_{\delta\gamma}
\ee
and the same relations for $ \Lhplusmin{\alpha}{\beta} $ with $ t $ and $ \hat
t $ exchanged.
These definition is extended to products by $ \Lplusmin {\alpha}{\beta}(ab) :=
\Lplusmin {\alpha}{\mu}(a)\Lplusmin {\mu}{\beta}(b)$, $ \forall a,b\in
SL_q(2,\ce) $, and the same for $ \Lhplusmin {\alpha}{\beta} $. The commutation
relations in $ \regfunlor $ and the Hopf algebra structure are a direct
consequence of
these definitions (cf. \cite{dswz}). The involution on the generators of
$ \regfunlor $ is $ {(\Lplusmin{\alpha}{\beta})}^{\dagger} =
S(\Lhplusmin{\alpha}{\beta}) $. Using the
properties of the $ SL_q(2) $-$ \RM $-matrix, some of these generators can be
eliminated, and it turns out that $ \regfunlor $ is generated by
$ \Lplus 11\; $,$ \;\Lplus 22\; $,$ ;\Lplus 12\;$,$\;\Lhplus 21\;$,
$\;\Lminus 11\;$,$\;\Lminus 12\;$,$\;\Lminus 21\;$,$\;\Lminus 22 $ with
$ \Lplus 22 = {(\Lplus 11)}^{-1} $. Moreover, in a
certain minimal extension also $ \Lminus 11 $ is invertible, so we are
essentially
left with six generators (cf. \cite{dswz}). We can now also define the algebra
of vector fields as the unital $ \ce $-algebra generated by the elements
$ \Yps {\alpha}{\beta} := \Lplus{\alpha}{\gamma}S(\Lminus{\gamma}{\beta})\in
\regfunlor $ and $ \Ypsh {\alpha}{\beta} :=
\Lhplus{\alpha}{\gamma}S(\Lhminus{\gamma}{\beta}\in
\regfunlor $. This algebra was introduced in \cite{cdswz,djswz}, and it was
shown that for $ SL_q(2,\ce) $ it is a sub-Hopf algebra of $ \regfunlor $ and
that a certain natural extension of the algebra of vector fields is isomorphic
to $ \regfunlor $. The left invariant vector fields are then given by
$ X := 1/\l({\bf 1} - Y) $. Let us finally mention that there exist two
Casimir operators $ C_1 $ and $ C_2 $ in $ \regfunlor $.

% kapitel 2 -------------------------------
\section{Angular Momentum Representation of $ \regfunlor $}

\subsection{Angular Momentum Algebra}
In the sequel we will always assume $ q\geq 1 $.\\
In the undeformed case we have a representation of the Lie algebra of
$ SL(2,\ce) $ in terms of antisymmetric generalized angular momentum
operators acting on functions over Minkowski space. In this section, we will
define the analogue of this in the q-deformed framework, i. e. a representation
of the quantum universal
enveloping algebra of $ SL_q(2,\ce) $ (which is essentially given by
$ \regfunlor $) by differential operators on quantum Minkowski space $ \mq $.
In \cite{gow} such a representation was found for the closely related case of
$ SO_q(N) $ and, apart from the star structure, these results can be applied to
the case of $ SL_q(2,\ce) $. Therefore we first introduce the elements
$ u^{ij}\in \difflor $ as
\be\label{defuij}
u^{ij} := (1+q^{-4})\Ci ij + q^{-4}(\qi\bfRmvier ijkl x^k\dobb l -q\bfRmivier
ijkl x^k\dobb l)
\ee
and define
\bea
V^{ij} & := & \mu\paind ijkl u^{kl} = q^{-4}\l\q2\paind ijkl \mu x^k
\dobb l,\\
U & := & \mu\C ij u^{ij} = (1+q^{-4})\mu({(q+\qi)}^2{\bf 1} +
(q^{-4} -1)\C ij x^i\dobb j).
\eea
These differential operators commute with $ r^2 $ and we are led to the
following
\begin{defin}
The $ \ast $-subalgebra $ \Uhutlor := <(V^{ij})_{i,j =
1,\ldots,4},U>\subset \difflor $ with involution $ \ast $ given by
\be\label{sternulor}
{(V^{ij})}^{\ast} = V^{kl}\C km\C ln\eta^n{}_i\eta^m{}_j,\qquad\qquad
U^{\ast} = U
\ee
is called {\em angular momentum algebra} on quantum Minkowski
space.
\end{defin}
As the rank of the antisymmetric projector in four dimensions is six, only six
of the generators $ V^{ij} $ are linearly independent.
To determine the action of the elements of $ \Uhutlor $ as differential
operators on $ \mq $, we consider the algebra
\be
\Ahutlor := <{\bf 1},(V^{ij}),U,(x^j)>_{i,j = 1,\ldots,4}\subset\difflor .
\ee
Using the relations in $ \difflor $, we can derive the relations in $ \Ahutlor
$,
which determines the action of the generators of $ \Uhutlor $ as differential
operators. The result is (cf. \cite{gow}):
\bea
V^{ij}x^k & = & -\q2\paind ijmn x^m V^{nk}+\frac{q\l}{(1+q^4)\q2}\paind ijmn
\Ci nk x^m U,\label{wivko}\\
U x^k & = & (q^4-q^{-4})\left(\frac{1}{\l{\q2}^2}x^k U -\frac{q}{2}\C mn
x^mV^{nk}
\right).\label{wiuko}
\eea
Furthermore the relations in $ \mq $ imply the following identities:
\begin{eqnarray}
0 & = & x^3V^{12} +x^2V^{31}+\l x^2 V^{12} -\frac{\l}{\q2} x^1
V^{14}-\frac{2\qi}
      {\q2} x^1 V^{23},\nonumber\\
0 & = &
x^4V^{12}+q^{-2}x^1V^{24}-\frac{\qi}{\q2}(q^2+q^{-2})x^2V^{14}-\frac{q^{-2}\l}
      {\q2} x^2V^{23}, \nonumber\\
0 & = & x^4 V^{31} +x^1V^{43} +\l
x^1V^{24}-\frac{q^2{\l}^2}{\q2}x^2V^{14}\label{xvrel}\\
  &   & -\,\frac{2q\l}{\q2} x^2V^{23} +\frac{2q}{\q2}x^3V^{14}-\frac{\l}{\q2}
x^3
      V^{23},\nonumber\\
0 & = & x^3V^{24}+q^{-2}x^2V^{43}-\frac{q^2\l}{\q2} x^4V^{14}-
\frac{2q}{\q2}x^4
      V^{23}.\nonumber
\end{eqnarray}
For explicit calculation a more convenient choice for the generators is:
\begin{equation}
\begin{array}{rcl}
\vepl & := & \frac{q^2}{\l}V^{12},\\
\vemi & := & \frac{q^2}{\l}V^{43},\\
M_1 & := & \frac{q^2}{\q2}(-V^{14} +V^{23}+q^{-2}C),\\
\vzpl & := & \frac{q^2}{\l}V^{31},\\
\vzmi & := & \frac{q^2}{\l}V^{24},\\
M_2 & := & \frac{q^2}{\q2}(-q^2V^{14} -V^{23}+q^{-2}C),
\end{array}
\end{equation}
where $ C $ is defined by $ C := 1/((1+q^{-4})\q2) U $.
For the commutators of the generators of $ \Uhutlor $ with each other we have
the identities (cf. \cite{gow})
\be
\paind ijkl\C mn V^{km}V^{nl} = \frac{\qi\l}{(q^2+q^{-2}){\q2}^2}V^{ij}U,
\qquad\qquad V^{ij}U = UV^{ij}\label{reluntalg}
\ee
and
\be\label{relcas}
{\bf 1}  =  \frac{1}{{(1+q^{-4})}^2{\q2}^4}U^2 -\frac{q^6}{2{\q2}^2}\C ij\C kl
V^{ik}V^{lj}.
\ee
The second equation yields two identities which read with $ i=1,2 $
\be
{\bf 1} + M_i^2 -\frac{2}{\q2}M_iC +\frac{q{\l}^2}{\q2}(V^-_i V^+_i +q^{-2}
V^+_i V^-_i) = 0,\label{reluntcas1}
\ee
and (\ref{reluntalg}) gives another six relations (i=1,2):
\bea
V^{\pm}_i M_i & = & q^{-2} M_i V^{\pm}_i,\nonumber\\
\qi V^-_i V^-_i -q V^-_i V^+_i & = & \frac{1}{\l}({\bf 1} - M_i^2).
\label{crulor1}
\eea
Note that the algebras generated by $ \{ \vepl,\vemi,M_1 \} $ or $ \{ \vzpl,
\vzmi,M_2 \} $ alone are isomorphic to the algebra of left invariant vector
fields on the quantum group $ SU_q(2) $ as defined in \cite{wdok}.
This is exactly analogous to the undeformed case, but unlike the classical
case, these two subalgebras do not commute:
\bea
\vepl\vzmi & = & q^2\vzmi\vepl,\nonumber\\
\vepl\vzpl & = & q^{-2}\vzpl\vepl,\nonumber\\
\vemi\vzmi & = & q^{-2}\vzmi\vemi,\nonumber\\
\vzmi M_1 & = & M_1\vzmi,\nonumber\\
\vepl M_2 & = & M_2\vepl,\label{crulor2}\\
M_2M_1 -M_1M_2 & = & {\l}^3\vepl\vzmi,\nonumber\\
M_1\vzpl -\vzpl M_1 & = & q\l\vepl (\q2 M_2 - C),\nonumber\\
M_2\vemi -\vemi M_2 & = & q\l\vzmi (C -\q2 M_1),\nonumber\\
\vzpl\vemi -q^{-2}\vemi\vzpl & = & \frac{\q2}{\l}(qM_2M_1
+\qi M_1M_2\nonumber\\
&& -(M_1+M_2)C +\frac{1}{\q2}C^2).\nonumber
\eea
(\ref{reluntalg}) means that $ C $ is central $ \Uhutlor $. Finally, the
commutation relations of
the scaling operator $ \mu $ with the coordinates imply that it commutes
with all the generators
of $ \Uhutlor $. The extension of $ \Ahutlor $ by $ \mu $ and $ \mu^{-1} $ will
be denoted by $ \Ahutlormu $. From (\ref{sternulor}) we obtain the involution
on the generators as $ \as{(\vepl)} = -q\vzmi $, $
\as{(\vemi)} = -\qi\vzpl $, $ \as{(M_1)} = M_2 $ and $ \as{C} = C $. Actually
there
is a natural extension of the Algebra $ \Uhutlor $.
We define the element $ H\in \Uhutlor $ by
\be
H := M_1M_2 -\qi\l^2\vepl\vzmi.\label{defh}
\ee
$ H $ satisfies
\be
H (x^1,x^2,x^3,x^4) = (q^{-2} x^1,x^2,x^3,q^2 x^4) H,\label{wihco}
\ee
which together with $ H({\bf 1}) = {\bf 1} $ implies that $ H $ is a positive
and invertible differential operator
on $ \mq $. Therefore the operator $ H^{1/2} $ and its inverse are well defined
by their commutation relations with the coordinates following from
(\ref{wihco}). The extension of $ \Uhutlor $ by $ H^{\pm1/2} $ will be denoted
by $ \Ulor $.

\subsection{Differential Representation of $ \regfunlor $}
In the undeformed case the algebra of angular momentum operators on Minkowski
space is a representation of the Lie algebra of $ SL(2,\ce) $ ,
i.e. a representation of the left invariant vector fields on the Lie group. It
turns out that a similar statement is true in the deformed case. To see this,
we
consider an algebra representation $ \pi:\regfunlor\ra\mbox{End}_{\scriptC}
(\mq), \quad a\mapsto \pi(a) =: a_{\pi},$ of $ \regfunlor $ by linear
operators on $ \mq $.
$ a_{\pi} $ is defined for an arbitrary $ f\in\mq $ by
\be
a_{\pi}(f) = (S^{-1}(a)\ot id)\circ\delta(f),
\ee
where $ \delta $ is the $ SO_q(3,1) $-coaction defined in section 1. $ U_{\pi}
:= \pi(\regfunlor) $ is a subalgebra of $ \mbox{End}_{\scriptC}(\mq) $. It
turns
out that $ U_{\pi} = \Ulor $ and we obtain the
\begin{prop}
The algebra $ \Ulor\subset\mbox{End}_{\scriptC}(\mq)  $ is a $ \ast $-
representation of the algebra $ \regfunlor $. In this representation the two
Casimir operators of $ \regfunlor $ coincide, i.e. $ \pi(C_1) =
\pi(C_2) = C $.
\end{prop}
In particular, it turns out that $ \pi $ maps the algebra of
vector fields on $ SL_q(2,\ce) $ to $ \Uhutlor $.
However, $ \mq $ being a comodule algebra, $ U_{\pi} $ and therefore $ \Ulor $
is also a bialgebra and even a Hopf algebra. Defining the mappings $ \co:
U_{\pi}\ra U_{\pi}\ot U_{\pi} $,$ \coe: U_{\pi}\ra\ce $ and $ S_{\pi}:
U_{\pi}\ra U_{\pi} $ by
\bea
(\co_{\pi}(a_{\pi}))(f\ot g) & := & a_{\pi}(fg) \qquad\forall\;f,g\in\mq,
\nonumber\\
\coe_{\pi}(a_{\pi}) & := & i(a_{\pi}({\bf 1})),\label{coproduktdarst}\\
S_{\pi}(a_{\pi}) & := & \pi(S^{-1}(a)),\nonumber
\eea
where $ i:\mq\ra\ce $ projects the subalgebra of $ \mq $ generated by $ {\bf 1}
\in\mq $ onto the complex numbers, we deduce $ \forall \;a\in \regfunlor: $ $
\co_{\pi}(a_{\pi}) = (\pi\ot\pi)(\tau\circ\co(a)) $, $
\coe_{\pi}(a_{\pi}) = a({\bf 1}), $ where $ \tau $ denotes the transposition
of tensor components and $ \co $
the coproduct in $\regfunlor $, and get the following
\begin{prop}\label{hopfstrukturulor}
$ U_{\pi} $ together with the coproduct $ \co_{\pi} $, the counit $ \coe_{\pi}
$,
the antipode $ S_{\pi} $ and the involution $ \as{a_{\pi}} := \pi(a^{\dagger})
$
is a $ \ast $-Hopf algebra.
\end{prop}

$ \Ulor $ contains a $ \ast $-subalgebra corresponding to vector fields on
$ SU_{q^{-1}}(2) $. It is generated by $ H^{\pm 1/2}, X^{\pm} $ with
\bea
X^+ & := & {\left(\frac{q}{\q2}\right)}^{1/2}(q\vepl C -q^2\vepl M_2 -\qi\vzpl
M_1),\label{defxplus}\\
X^- & := &  {\left(\frac{\qi}{\q2}\right)}^{1/2}(-\vzmi C +\qi\vzmi M_1
-\qi\vemi
M_1)\label{defxminus}
\eea
and involution $ \as H = H $ and $ \as{(X^+)} = qX^-\, $. It will be denoted
$ \Udreh $, its extension by the generators of $ \mq $ by $ \Adreh $.
The relations in $ \Udreh $ are those already
encountered in (\ref{crulor1}). The element $ W\in\Udreh $, related to the
central element $ S^2\in U_{q^{-1}}(SU(2)) $ (cf. \cite{wdok}) by
$ W:= q\pi(\l^2S^2 + \q2{\bf 1}) $ is the central Casimir operator of
$ \Udreh $. It satisfies
\be
HW = q^2{\bf 1} + H^2 + q^2\l^2\q2 X^-X^+,\qquad\qquad\qquad\as W= W.
\ee
Finally, we give the relations determining $ H,X^{\pm} $ and $ W $ as operators
on $ \mq $. They read
\bea
H(t,y,z,x) & = & (t,q^{-2}y,z,q^2x)H,\nonumber\\
X^+(t,y,z,x) & = & (t,y,z,x)X^+ + (0,0,\qi y,-z)H\nonumber\\
X^-(t,y,z,x) & = & (t,y,z,x)X^- + (0,z,-qx,0)H\label{crkodreh}\\
W(t,y,x,z) & = & (t,q^2y,z,q^{-2}x)W +\l\q2 (0,-y,q\l z,q\l x)H\nonumber\\
           &   & + \l^2\q2 [(0,q^2 z,-qx,0)X^+ +(0,0,qy,-z)X^-].
\eea
As $ H $ and $ X^{\pm} $ are a representation of vector fields of the
q-deformed
rotation group, they commute with the generator $ t\in\mq $ corresponding to
the time coordinate in the limit $ q\ra 1 $.

\section{Algebra of Observables}
We want to define momentum coordinates
$ p^i $, such that the algebra generated by the momenta is isomorhpic to
$ \mq $ as a $ \ast $-comodule algebra of $ SO_q(3,1) $. Unlike the undeformed
case, the partial derivatives $ \dob i $ are not closed under involution.
\begin{lem}
The momentum operators $ (p^j)_{j=1,\ldots,4}\in\difflor $ defined by
\be
p^j := -i(\dob j +q^{-4}\dobb j)
\ee
generate the momentum quantum space $ \mqp $ of functions over q-Minkowski
space, i.e. they satisfy
\be
\paind ijkl p^kp^l = 0,\qquad\qquad\as{(p^j)} = p^k\C kl\eta^l{}_j.
\ee
\end{lem}
\begin{sloppypar}
The complete symmetry between coordinates and momenta as $ \ast $-comodule
algebras of $ SO_q(3,1) $ is established by the following\end{sloppypar}
\begin{lem}
The algebra $ \Aplor $ generated by $ (p^i) $ and $ \Ulor $ is isomorphic
to $ \Alor $ with the isomorpism $ i $ given on the generators by $ i(p^i) =
x^i $,$ i(V^{ij}) = V^{ij} $ and $ i(U) = U $.
\end{lem}
Now we are in the position to introduce the algebra of observables (cf.
\cite{gow}) by the following
\begin{defin}
The algebra of observables $ {\cal O} := <{\bf 1},(x^i)_{i=1,\ldots,4},
(p^j)_{j=1,\ldots,4}> $$\subset\difflor $ is the algebra
generated by coordinates and momenta.
\end{defin}
\begin{prop}
In $ {\cal O} $ we have relations
\be
p^ix^j -q\Rmi ijkl x^kp^l = -iu^{ij}\label{qccr}
\ee
where $ u^{ij}\in\difflor $ are the elements defined in (\ref{defuij}).
\end{prop}

The next aim is to
construct a Hilbert space representation of this algebra of observables, which
can be interpreted as the state space for q-deformed
relativistic quantum mechanics. To make the connection with the angular
momentum algebra, we rewrite (\ref{qccr}) as
\be
p^ix^j - q\Rmi ijkl x^kp^l = -i\mu^{-1}(V^{ij}+U).
\ee
It was shown in \cite{gow} that we can reconstruct the partial derivatives
given the coordinates and the angular momentum algebra with the help of the
following identities in $ \Alormu $:
\bea
r^2\dob i & = & \frac{1+q^{-2}}{1-q^2}\,x^i +\frac{1}{q^8 -1}\,x^iU\mu
+\frac{1+q^{-2}}{2(q^2-1)}\,\C klx^kV^{li}\mu,\\
r^2\dobb i & = & -\frac{1+q^2}{q^{-2}-1}\,x^i +\frac{1}{q^{-8}
-1}\,x^iU\mu^{-1}
+q^2\frac{1+q^2}{2(1-q^{-2})}\,\C klx^kV^{li}\mu^{-1}.
\eea
Due to the complete symmetry between coordinates and momenta, the same
relations
hold with coordinates and momenta exchanged. This means that we can find
Hilbert space representations $ {\cal O} $ by
constructing Hilbert space representations of $ \Alormu $ with invertible
$r^2 $. This will be done in the next section.

%kap3--------------------------------------------------------

\section{Hilbert Space Representation of $ {\cal O} $}
\subsection{Representation of $ \Apdreh $}
In analogy to the undeformed case we expect a relativistic one particle state
to be an element of an
irreducible representation of the Poincar\'{e} algebra. As a maximal set of
commuting observables we take the square of the momentum $ p^2 := \C ij p^ip^j
$
which is the square of the particle mass, the energy $ p^0 $ (corresponding
to the generator t in the coordinate algebra), the angular
momentum $ W $ and the 3-component of angular momentum $ H $. In the limit
$ q\ra 1 $ this is a complete set of commuting observables for spin 0. We will
construct an irreducible $ \ast $-representation of $ \Aplor $ as a direct sum
of irreducible representations of $ \Apdreh $.

The hermitean elements $ p^2\,,\,p^0 $ are central in $ \Apdreh $. In an
irreducible representation we can therefore choose them as multiples of the
identity operator. We define for abitrary $ M^2\,,\,E\in\R $ a linear space
\be
{\bf H}^{M,E} := \mbox{Lin}\{a\ket ME00 :a\in\Apdreh\},
\ee
where the cyclic vector $ \ket ME00 $ has the properties
\bea
X^{\pm}\ket ME00 := 0, & {} & H\ket ME00 := \ket ME00.\label{widrehskalar}\\
p^2\ket ME00 := M^2\ket ME00, & {} & p^0\ket ME00 := E\ket
ME00,\label{wirtskalar}
\eea
i.e. it carries a scalar representation of $ \Udreh $.
\begin{sloppypar}The commutators (\ref{crkodreh}) imply $ {\bf H}^{M,E} =
Lin\{a\ket ME00:a
\in\mqp\} $ . Considering the vector $ p_y^l\ket ME00\in {\bf H}^{M,E} $ with
$ l\in \N_0 $, we find it to be an eigenvector of $ H $ and $ W $ with
eigenvalue $ l $ for both operators which is annihilated by $ X^+ $. Therefore
it is a highest weight for the irreducible representation of $ U_q(SU(2)) $
characterized by the eigenvalue l of the Casimir $ W $.\end{sloppypar}
If we define $ \gamma_{E,l}\ket MEll := p_y^l\ket ME00 $,
then a basis of this representation is given by the states $ \ket MElm $ with
$ m = -l,\ldots,l $ defined by
\be
\ket MElm := q^{m+1}{\left(\frac{\qzahl
2q}{\qzahl{l+m+1}q\qzahl{l-m}q}\right)}^{-1/2}
X^-\ket MEl{m+1}.
\ee
Being eigenvectors to different eigenvalues of $ H $, these states are linearly
independent.
\begin{lem}
The set $ B :=\{\ket MElm $ mit $ l\in\N_0,\;m\in \Z,\;|m|\leq l \} $
is a linear basis of $ {\bf H}^{M,E}. $
\end{lem}
For $ \bra ME{l^{\prime}}{m^{\prime}}\cket MElm := \delta_{l^{\prime}l}
\delta_{m^{\prime} m} $ to be a consistent definition of a scalar product
on $ {\bf H}^{M,E} $, the constant $ \gamma_{E,l} =
\bra ME00 {(\as y)}^l y^l\ket ME00 $ must be positive.
Using the definition of the central element $ p^2 $ and the relations
(\ref{crkodreh}), we get a recursion formula for $ \gamma_{E,l} $:
\be
\gamma_{E,l+1} = \frac{q^2\qzahl{l+1}q}{\qzahl{2l+3}q}
(\frac{1}{\aqzahl{l+1}q}t^2-\aqzahl{l+1}qM^2)\gamma_{E,l}
\ee
with $ \aqzahl xq := \frac{q^x+q^{-x}}{q+\qi} $.
As $ \gamma_{E,0}=1 $ this fixes $ \gamma_{E,l} $ for all $ l\in\N_0 $.
We have to distinguish two cases:
\bea
i)\quad M^2\leq 0 & : & \quad \gamma_{E,l}>0\quad \forall\,l\in\N_0. \\
ii)\:\; M^2>0     & : & \gamma_{E,l}\stackrel{l\ra\infty}{\ra} -\infty.
\eea
In the first case there is no restriction on the values of $ E $, but for the
physically interesting case of real mass the requirement of positivity forces
the recursion to terminate, i.e. for each fixed value of $ M^2>0 $ there must
be an
$ N\in\N_0 $, such that $ \gamma_{E,l} = 0\quad\forall\;l>N $.
This means that in this case we get only the discrete energy eigenvalues
\be\label{tew}
E=\pm\aqzahl{N+1}q M.
\ee
In the sequel we take $ M^2>0 $. We can then use the positive integer $ N $
to label the states by $ \ket MNlm $.
The sign of the energy eigenvalues is an additional invariant of the
representation, because all the generators of $ \Apdreh $ commute with $ p^0 $.
So we have two orthonormal bases which span the linear spaces
\be
{\bf H}_{\pm}^{M,N} := \mbox{Lin}\{\ketpm MNlm :
N,l\in\N_0\;,l\leq N\;,m\in z\;|m|\leq l\}.
\ee
Completing $ {\bf H}_{\pm}^{M,N} $ to Hilbert spaces $ \h_{\pm}^{M,N} $ with
respect
to the scalar product defined above, we get
\begin{lem}
For every $ M^2 > 0 $ and $ N\in\N_0 $ the Hilbert spaces $ \h_{\pm}^{M,N} $
are irreducible $ \ast $-
representations of the algebra $ \Apdreh $.
\end{lem}

\subsection{Representation of $ \Aplor $ }
We consider now the linear space
\be
{\bf H}_{\pm}^M := \bigoplus_{N=0}^{\infty} {\bf H}_{\pm}^{M,N}
\ee
with scalar product $ \brapm M{N^{\prime}}{l^{\prime}}{m^{\prime}}\cketpm MNlm
:=
\delta_{N^{\prime}N}\delta_{l^{\prime}l}\delta_{m^{\prime}m} $.
To show that these spaces are representations of $ \Aplor $, it is sufficient
to determine the action of the generators of $ \Ulor $ on the $ \Udreh
$-invariant
vectors $ \ketpm MN00 $, because the action on an arbitrary vector $ \ketpm
MNlm $
can then be deduced using the relations in $ \Aplor $. Making use of the
commutators of the generators of $ \Ulor $ and $ \Udreh $ and of the
discreteness
of the spectrum of $ p^0 $, we make the following ansatz:
\bea
V^+_k\ketpm MN00 & = & \sum_{i=0}^{\infty}c_{\pm,N,k,i}^+\ketpm
Mi11,\nonumber\\
V^-_k\ketpm MN00 & = & \sum_{i=0}^{\infty}c_{\pm,N,k,i}^-\ketpm
Mi1{-1},\label{ansatz}\\
M_k\ketpm MN00 & = & \sum_{i=0}^{\infty}c_{\pm,N,k,i}\ketpm MN10 +
c_{\pm,N,k,i}^{\prime}\ketpm MN00,\nonumber
\eea
with $ k = 1,2 $. This also determines the action of the Casimir $ C $, because
one
can prove $ C\ketpm MN00 =(qM_1 +\qi M_2)\ketpm MN00 $.
The commutation relations of the generators of $ \Ulor $ with $ X^+ $ and
$ X^- $ can be used to eliminate all but two coefficients:
\begin{equation}
\begin{array}{rcl}
c_{\pm,N,1,i}^{\prime} & = & c_{\pm,N,2,i}^{\prime}\,,\\
c_{\pm,N,1,i}^+ & = & {(q\q2)}^{1/2}\frac{1}{\l}\;c_{\pm,N,1,i}\,,\\
c_{\pm,N,1,i}^- & = & -{(\qi\q2)}^{1/2}\frac{1}{\l}\;c_{\pm,N,1,i}\,,\\
c_{\pm,N,2,i}  & = & -q^2\;c_{\pm,N,1,i}\label{relkoeff1}\,,\\
c_{\pm,N,2,i}^+ & = & q^3{(q\q2)}^{1/2}\frac{1}{\l}\;c_{\pm,N,1,i}\,,\\
c_{\pm,N,2,i}^- & = & -{(q\q2)}^{1/2}\frac{1}{\l}\;c_{\pm,N,1,i}\,.
\end{array}
\end{equation}
Dropping the index 1 we denote the remaining coefficients by $ c_{\pm,N,i} $
and $ c_{\pm,N,i}^{\prime} $. The commutation relations of the generators of
$ \Ulor $ with $ p^0 $ can be used to determine the possible values of $ i $.
For $ i\geq 1 $ we find
\be\label{soli}
i = N\pm1.
\ee
Furthermore we get the result that $ i=0 $ can only appear for $ N=1 $ which
is consistent
with (\ref{soli}).
Moreover we get for $ i \geq 1 $ relations between $ \csind Ni $ and $ \cind Ni
$.
Finally, using these relations, the identity $
H\ketpm MN00 = (M_1M_2-\qi\l^2\vepl\vzmi)\ketpm MN00 = \ketpm MN00 $ and the
fact that
we want a $ \ast $-representation of $ \Aplor $ on $ {\bf H}^M $, we can
determine all the coefficients. Choosing a common
phase to make the coefficients real, we obtain as the final result:
\bea
\csind N{N+1} & = & \csind{N+1}N = \frac{1}{\q2}\quad\mbox{f\"ur}\quad N\geq
0,\\
\csind N{N-1} & = & \csind{N-1}N = \frac{1}{\q2}\quad\mbox{f\"ur}\quad N\geq
1,\\
\cind N{N+1} & = & \mp\frac{\qi}{\q2}{\left(\frac{\qzahl{N+3}q}{\qzahl
3q\qzahl{N+1}q}
\right)}^{1/2}\quad\mbox{f\"ur}\quad N\geq 0,\\
\cind N{N-1} & = & \pm\frac{\qi}{\q2}{\left(\frac{\qzahl{N-1}q}{\qzahl
3q\qzahl{N+1}q}
\right)}^{1/2}\quad\mbox{f\"ur}\quad N\geq 1.
\eea
Checking now the consistency of this solution with all the relations in $
\Aplor $,
and completing $ {\bf H}_{\pm} $ to the Hilbert space $ \h_{\pm}^M $ we get
\begin{sloppypar}
\begin{prop}
The Hilbert space $ \h_{\pm}^M $ is an irreducible $ \ast $-representation of
the
algebra $ \Aplor $.
\end{prop}
\end{sloppypar}
This is in particular a representation of the q-deformed Poincar\'{e} algebra
(cf. \cite{psw}).

\subsection{Representation of $ {\cal O} $}
For fixed $ M^2>0 $ the operator $ p^2 $ is constant and postive on
$ {\bf H}_{\pm}^M $ and the operators $ p := {(p^2)}^{1/2} $,$  p^{-1}
:= {(p^2)}^{-1/2} $ and $ p^0/p $ with $ p^0/p \ketpm MNlm = \pm\aqzahl{N+1}q $
are well-defined. $ \mu $ does not commute with $ p^2 $, so it connects
different irreducible representations.

Let $ M_0>0 $ be fixed, $ M_k := M_0 q^k $ with $ k\in\Z $ and
\be
{\bf H}_{\pm} := \bigoplus_{k}{\bf H}^{M_k}_{\pm}
\ee
with scalar product $ \brapm
{M_{k^{\prime}}}{N^{\prime}}{l^{\prime}}{m^{\prime}}
\cketpm {M_k}Nlm :=
\delta_{k^{\prime}k}\delta_{N^{\prime}N}\delta_{l^{\prime}l}
\delta_{m^{\prime}m} $.
Denoting the Hilbert space completion of $ {\bf H}_{\pm} $ by $ \h_{\pm} $, we
have
\begin{prop}\sloppy
$ \h_{\pm} $ is an irreducible $ \ast $-representation of $ \Aplormu $ with $
\mu $
acting as $ \mu\ketpm{M_k}Nlm = \ketpm {M_{k-1}}Nlm $. $ p^2 $ is invertible on
$ \h_{\pm} $, so this is in particular a representation of the algebra
of observables $ {\cal O} $.
\end{prop}

In this paper, we have found a Hilbert space representation for a deformation
of the Heisenberg algebra. In quantum mechanics the operators $ (x^i) $ and
$ (p^i) $ must be observables, i.e. their restriction to a common domain must
be essentially self-adjoint. The operators $ p^2 $ and $ p^0 $, being diagonal
in this representation, are already essentially self-adjoint on the Hilbert
space. This implies also that all the momenta have this property. The question
however is, whether $ r^2 $ and $ t $ are essentially self-adjoint. To solve
this problem, we
have to diagonalize $ {{r^2}}^{\dagger} $. The transformation between the
eigenvectors
of $ p^2 $ and $ r^2 $ is then a generalized Fourier transformation. With the
help of the operators $ (x^i) $ one could then try to discuss the problem
of locality.

% kapitel 4 -------------------------------

% ------------------------------------------------------------
% Anhang
%\begin{appendix}
%\include{a:applor}
%\include{a:appcrvektor}
%\include{a:appcrlor}
%\end{appendix}
% ------------------------------------------------------------
% Literaturverzeichnis

% ------------------------------------------------------------
\end{document}